\documentclass[twocolumn,10pt]{iopart}

\usepackage{epsfig}
\usepackage{graphicx,color}

\usepackage{iopams}

\usepackage{graphicx,color}
\usepackage{epsfig}

\usepackage{dcolumn}

\usepackage{bm}
\usepackage{amssymb}
\usepackage{array}

    \newcommand{\ket}[1]{\ensuremath{|\,{#1}\,\rangle}}
    \newcommand{\bra}[1]{\ensuremath{\langle\,{#1}\,|}}

    \newcommand{\lsub}[1]{\ensuremath{_{_{\!\scriptstyle #1}}}}
    \newcommand{\ce}[1]{\ensuremath{\mathcal{#1}}}
    
    \newcommand{\st}{\scriptstyle}

    \renewcommand{\vec}[1]{\mbox{\boldmath{\ensuremath{{#1}}}}}

    \newcommand{\itg}[1]{\ensuremath{\int\!\!d{#1}\!\!}}
    \newcommand{\itgf}[1]{\ensuremath{\int\!\!d{#1}\,}}

    \newcommand{\sinc}{\ensuremath{\mbox{\hspace{1.3pt}sinc}\,}}

\begin{document}

\title{Measurement of spatial qubits}

\author{G Lima$^{1,2}$, F A Torres-Ruiz$^2$, Leonardo Neves$^1$,  A Delgado$^2$, C Saavedra$^2$ and S P\'adua$^1$}

\address{$^1$ Departamento de F\'{\i}sica,
Universidade Federal de Minas Gerais, Caixa Postal 702, Belo~Horizonte, MG 30123-970, Brazil.}

\address{$^2$ Center for Quantum Optics and Quantum Information,
Departamento de F\'{\i}sica, Universidad de Concepci\'on,  Casilla 160-C, Concepci\'on, Chile.}

\ead{glima@udec.cl}

\begin{abstract}

In this paper we study the state determination for composite systems of two spatial qubits. We
show, theoretically, that one can use the technique of quantum tomography to reconstruct the density matrixes of these systems. This tomographic reconstruction is based on the free evolution of the spatial qubits and a postelection process.

\end{abstract}



\submitto{\jpb}
\maketitle

\section{Introduction}
\label{sec:intro}

The concept of quantum state plays a central role in Quantum Theory. The statistical distributions for the results of experiments carried out on a physical system can be completely predicted from its initial state. This has led to the development of techniques for the state determination. In the field of atomic physics, quantum endoscopy was used to determine the state of ions and atoms \cite{Vogel1,Schleich,Walmsley}. In quantum optics, the Wigner function of multimode fields could be measured using homodyne detection \cite{Vogel2,Smithey,Vogel3}, and quantum tomographic reconstruction (QTR) was used for measuring the polarization state of the parametric down-converted photons \cite{Kwiat,James,White,Martini,Kwiat2}. Beside this, the raising of new technologic fields has brought more motivations for the study of these techniques of reconstruction. For example, in the field of Quantum Information, protocols like teleportation \cite{Bennett} and superdense coding \cite{Bennett2} require initially knowing the quality of the quantum channel which is being used for implementing them.

In spontaneous parametric down-conversion (SPDC) one photon from a pump beam incident onto a non-linear crystal originates, with small probability, two other photons usually called signal (s) and idler (i) or also twin photons \cite{MandelBook}. Recently, we have demonstrated that by exploring the transverse momentum correlations of the twin photons it is possible to generate entangled states of two effective $D$-dimensional quantum systems \cite{Leonardo,GLima}. Because the dimension of the Hilbert space of these photons is defined by the number of different ways available for their transmission through apertures where they are sent to, we call them spatial qudits.

In this work we investigate the state determination of these quantum systems. We show, theoretically, that one can implement the process of QTR to obtain the density operators of two spatial qubits ($D = 2$) states. The theoretical description of our protocol is closely related to the description given in Ref.~\cite{James} for the quantum tomography of polarized two-qubit states. Even though we have considered just the case of qubits, the description used here can be generalized to spatial qudits states.

\section{The state of two spatial qubits}
\label{SECPSI}

The angular spectrum of the pump beam is transferred to the two-photon state generated by SPDC,
when the monochromatic, paraxial and thin crystal approximations are assumed \cite{Monken,Gera1,Rubin,Sergi,Edu1}.
The spectral matching is implemented by inserting narrow bandwidth interference
filters in front of the detectors \cite{Monken,Edu1}. The state of twin photons when they are transmitted through generic apertures, under above approximations, can be written in one dimension as \cite{Leonardo,Monken}

\begin{equation}   \label{Psi}
\ket{\Psi}  =  \itg{q_{s}}\itgf{q_{i}}\ce{F}(q_{s},q_{i})
\ket{1q_{s}}\ket{1q_{i}},
\end{equation} where $\ket{1q_{j}}$ is the Fock state for one
photon in mode $j= s,i$ with transverse wavevector $q_{j}$. $\ce{F}(q_{s},q_{i})$ is
the two-photon amplitude and it is given by
\begin{eqnarray}
\ce{F}(q_{s},q_{i}) & =  &
\itg{x_{s}}\itgf{x_{i}}  A_{s}(x_{s}) A_{i}(x_{i})
\exp{\bm{(}ik_p (x_{i}-x_{s})^{2}/8z_{A}\bm{)}}
 \nonumber \\ &  &  \times \;
  W\bm{(}{\st\frac{1}{2}}(x_{s}+x_{i});z_{A}\bm{)}
\exp{\bm{(}-i(q_{s}x_{s} + q_{i}x_{i})\bm{)}}.
\label{biphoton1}
\end{eqnarray}
The longitudinal distance $z_A$ defines the plane where the apertures are placed, and $k_p$ is the wave number of the pump beam used to generate the twin photons. The function $W(x;z_A)$ is the pump field distribution at the plane of the apertures ($z=z_{A}$) and at the transverse position $x$. $A_{j}(x_{j})$ is the transmission function of the aperture in mode $j$.

In the case where these apertures are two identical double slits, the two-photon state in Eq.~(\ref{Psi}) simplifies to \cite{Leonardo,GLima}
\begin{eqnarray}             \label{estqubit}
\ket{\Psi} & = & W_{+,+}\ket{+}\lsub{s}\ket{+}\lsub{i} +
               W_{+,-}\ket{+}\lsub{s}\ket{-}\lsub{i} \nonumber\\
           &   & + W_{-,+}\ket{-}\lsub{s}\ket{+}\lsub{i} +
               W_{-,-}\ket{-}\lsub{s}\ket{-}\lsub{i} ,
\end{eqnarray}
where $W_{+,-} = W_{-,+} \equiv \gamma\, W(0;z_{A})e^{ik_p d^{2}/8z_{A}}$  and  $W_{\pm ,\pm} \equiv \gamma\, W(\pm/2;z_{A})$. $d$ is the separation between two consecutive slits and $a$ is each slit's half width. $\gamma$ is a normalization constant. The $\ket{+}_j$ (or $\ket{-}_j$) state is a single-photon state defined, up to a global phase factor, by
\begin{equation}      \label{base}
\ket{\pm}\lsub{j} \equiv \sqrt{\frac{a}{\pi}} \itgf{q_{j}}
e^{\mp iq_{j}ld}\sinc(q_{j}a)\ket{1q_{j}},
\end{equation}
and represents the photon in mode $j$ transmitted by the upper (lower) slit of its respective double slit. These states form an orthonormal basis in the two-dimensional Hilbert space of each photon and are used to define the twin photons' logical spatial qubit states. Therefore, it is clear that the state $\ket{\Psi}$ represents a system composed of two spatial qubits.

\begin{figure}[tbh]
\begin{center}
\rotatebox{-90}{\includegraphics[width=0.25\textwidth]{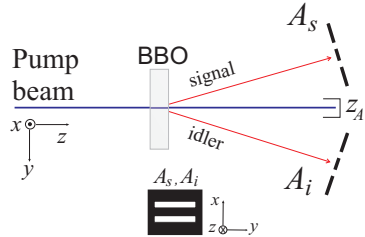}}
\end{center}
\caption{Schematic diagram of the experimental setup used for
generating spatial qubits. BBO is a non-linear crystal used to generate the
twin photons. $A_s$ and $A_i$ are the double-slits placed at
signal and idler propagation paths, respectively.} \label{Fig1}
\end{figure}

A typical setup used to generate the spatial qubits is outlined in Fig.~\ref{Fig1}. A pump beam incident upon a non-linear crystal generates SPDC the twin photons which are correlated in their transverse momenta. These photons are then sent through identical double slits ($A_i$ and $A_s$) placed at the $z_A$-plane and, after being transmitted by these apertures, the twin photons will be in the state of Eq.~(\ref{estqubit}).

\section{Detecting the spatial qubits in distinct bases}

For doing the quantum tomographic reconstruction of two-qubit states codified in polarization it is necessary to consider the detection of these qubits in some polarization states. The measurement bases that are usually considered in the experiments are \cite{James}: $\{| H \rangle,| V \rangle\}$ (horizontal and vertical polarizations), $\{| D \rangle,| A \rangle \}$ (linearly polarized light at $45^\circ$ and $-45^\circ$ with respect to the horizontal polarization, respectively), and $\{| L \rangle,| R \rangle \}$ (left and right circularly polarized lights, respectively). The measurement (the projection) in one of the states of these bases is represented by a projector operator, and an adequate choice among the available projections allows the reconstruction of the composite state.

Now we show how the spatial qubits can be measured in distinct bases of \{$\ket{\pm}_s,\ket{\pm}_i$\}. After being transmitted by the double slit, the twin photons freely evolve in space. Because of diffraction, it spreads out faster in the $x$-direction so that it passes from the discrete states \ket{\pm}, to states continuously distributed along this transverse direction. In a certain $z$-plane situated far behind the double slits' plane ($z=z_A$), each particular superposition of the basis states $\ket{\pm}$ will have a specific transverse probability distribution \cite{ConcEsp}. To recover the discrete nature of the logical states in this $z$-plane, we need to implement an adequate postelection process for the twin photons. As we shall show in the following lines, this can be properly done by allocating in this plane one double slit for the transmission of each mode $j$. We refer to these new apertures as detection double slits. The transverse position of their slits are $x^{\mu}_{j,k}$ where the subscript $j$ denotes the mode at which the double slit was placed (again $j=s,i$) and $k=0,1$ stands for the lower and upper slit of this aperture, respectively. The slits' width is $2b$. The superscript $\mu$ is used to define a set of measurements. For a certain value of $\mu$ we have two transverse positions available, $x^{\mu}_{j,0}$ and $x^{\mu}_{j,1}$, for the slits of the double slit. It can be seen as the mean position of the double slit's center and, as we shall see, it is the index which defines the spatial measurement basis which is being considered.  The state of the twin photons that crossed these two detection double slits can be written as
\begin{eqnarray}
|\,{\Phi }\rangle &\propto & \sum_{k,k'=0,1} B_{k,k'}|f(x^{\mu}_{s,k})\rangle |f(x^{\mu}_{i,k'}) \rangle,
\label{psitrans}
\end{eqnarray}
where the states $|f(x^{\mu}_{j ,k})\,\rangle$ are the states of the postselected photons which crossed the slit in the transverse position $x^{\mu}_{j ,k}$ of the double slit placed in mode $j$. These states are given by
\begin{eqnarray}
|f(x^{\mu}_{j ,k})\,\rangle &\equiv &\sqrt{\frac{b}{\pi }}\int
\!\!dq_{j}\,\exp(-iq_{j}x^{\mu}_{j ,k}) \nonumber \\
& &\times \mbox{sinc}\left( q_{j}\frac{x^{\mu}_{j,k}b}{2\alpha}
+q_{j}b\right) |1q_{j}\,\rangle , \label{fstates}
\end{eqnarray}
where $\alpha =(z-z_{A})/k_p$ and one can see that the logical states used to describe the twin photons are once again discrete. The calculation of the state $|\,{\Phi }\rangle$ is shown in \ref{A1}. However, what is really interesting to note now is that there is a relation between the coefficients of this state and the coefficients of the original
state of the twin photons given by  Eq.~(\ref{estqubit}). That is
\begin{eqnarray}
B_{k,k'}= \sum_{u,v=\pm} r_{u}(x^{\mu}_{s,k})r_{v}(x^{\mu}_{i,k'}) W_{u,v},
\label{rel}
\end{eqnarray}
with
\begin{equation}
r_{\pm }(x^{\mu}_{j ,k}) \equiv
\sinc\left(\frac{\left(x_{j,k}^{\mu}\mp d\right)a}{2\alpha}\right)
\exp\left(i\frac{\left(x^{\mu}_{j,k}\mp d\right) ^{2}}{4\alpha }\right).
\label{r}
\end{equation}
Therefore, one can think of the state of the twin photons transmitted through the detection double slits [Eq.~(\ref{psitrans})] as the original state of these photons [Eq.~(\ref{estqubit})] rewritten in the new basis \{$|f(x^{\mu}_{s,k})\,\rangle,|f(x^{\mu}_{i,k'})\,\rangle$\} (again $k,k'=0$ and $1$). As we mentioned before, the spatial bases are designed in terms of $\mu$. For distinct values of $\mu$ one has distinct transverse positions $x^{\mu}_{j,0}$ and $x^{\mu}_{j,1}$ for the slits of the detection double slit in mode $j$. The states $|f(x^{\mu}_{s,0})\,\rangle$ and $|f(x^{\mu}_{s,1})\,\rangle$ are orthornomal, i.e, they form an orthornomal basis for the Hilbert space of the single photon after the detection double slit. We use the superscripts $\mu,\mu',\mu'',...$ to denote distinct bases. The post-selection of the twin photons in one
of the basis states of \{$|f(x^{\mu}_{s,k})\,\rangle,|f(x^{\mu}_{i,k'})\,\rangle$\} is then represented by the following projector:
\begin{equation}
\textsl{P}_{x^{\mu}_{k},x^{\mu}_{k'}} =
\ket{f(x^{\mu}_{s,k}),f(x^{\mu}_{i,k'})}\bra{f(x^{\mu}_{s,k}),f(x^{\mu}_{i,k'})}.
\label{Pz}
\end{equation}

\section{Reconstructing the qubits state}

As is discussed in Ref \cite{James}, the density operator of two qubits is specified by sixteen real parameters. Thus, for determining their state, one needs to construct sixteen equations that are linearly independent and that are functions of measurable quantities. From now on we refer to these equations as MLI equations (measurable and linearly independent equations). In this section we show how this can be done for the spatial qubits state of Eq.~(\ref{estqubit}). To do this we first remind the reader of some basic results of the two photons interferometry theory. The photodetection of a $s$-photon (at $\vec{r}$,t) followed by the detection of a $i$-photon (at $\vec{r}'$,$t'$) is described by the fourth-order
correlation function which is defined as \cite{MandelBook}
\begin{equation}
G^{(2)}(\vec{r},\vec{r}',t,t')\! = \! \bra{\Phi}E^{(-)}_i(\vec{r},t)
E^{(-)}_s(\vec{r}',t')E^{(+)}_s(\vec{r}',t')E^{(+)}_i(\vec{r},t)\ket{\Phi},
\end{equation}
where $\ket{\Phi}$ is the two photon state and $E^{(\pm)}$ are the positive and negative parts of the electric field operator. The coincidence rate of these photons (at $\vec{r}$,t,$\vec{r}'$,$t'$) is, therefore, given by
\begin{equation}
C_{\vec{r},\vec{r}',t,t'} = \xi G^{(2)}(\vec{r},\vec{r}',t,t'), \label{C}
\end{equation}
where $\xi$ is a constant related to the total number of photon pairs sent to the detection apparatus and to its efficiency.

\subsection{The protocol}
\label{Prot}

We can now explain the protocol for doing the reconstruction of the two spatial qubits state given by Eq.~(\ref{estqubit}). To start with this, let us first assume the following general formula for its density operator $\rho$:
\begin{eqnarray}
\rho &\equiv &\ket{\Psi}\bra{\Psi} \nonumber \\
&= &\sum_{l,m,l',m'} \rho_{l,m;l',m'}\ket{l_s;m_i}\bra{l'_s;m'_i},
\end{eqnarray}
where the indices $l,m,l',m'$ assume the values $+$ (or $-$) when denoting the upper (lower) slits of the double slits $A_j$'s. For example, the matrix element, $\rho_{+,-;+,-}$, is the probability of detecting the twin photon signal at the slit ``+" and the idler photon at the slit ``-".

\begin{figure}[tbh]
\begin{center}
\includegraphics[width=0.5\textwidth]{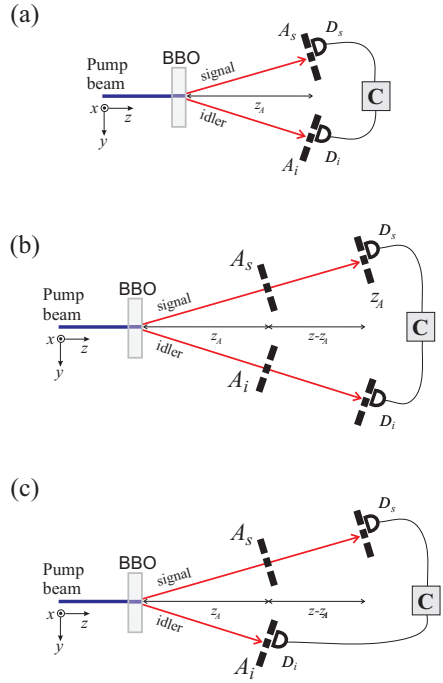}
\end{center}
\caption{Schematic diagram of the spatial measurements needed for doing the
 quantum tomography of two spatial qubits. In (a) the coincidence counts are
 recorded with the detectors ($D_s$ and $D_i$) fixed behind the slits of the
 double slits $A_s$ and $A_i$, respectively. In (b) they are fixed behind
 the detection double slits whose slit transverse positions are: $x^{\mu}_{j,0}=0$
 and $x^{\mu'}_{j,1}=\frac{\alpha \pi}{2d}$. In (c) one detector records the idler
 photons transmitted by the slits of $A_i$ and the other, the photons
 transmitted by the signal detection double slit.} \label{Fig2}
\end{figure}

To construct the first set of MLI equations, one should first measure the coincidence rates with two detectors ($D_i$ and $D_s$) placed just behind the slits of the apertures $A_i$ and $A_s$  as is shown in Fig~\ref{Fig2}(a) \footnote{For practical purposes, it is worth mentioning that this measurement can also be done in the plane of image formation of these apertures \cite{GLima2}.}. Four MLI equations can be constructed by permuting the detector positions behind the slits of these two double slits. The coincidences rates behind these slits, according to  Eq.~(\ref{C}), can be written as
\begin{eqnarray}
C_{l,m} &= & \eta N Tr(\rho \textsl{P}_{l,m} ),
\end{eqnarray}
 where $N$ is the number of the photon pairs transmitted through the double slits and $\eta$ the coincidence counts efficiency for the detectors used. $\textsl{P}_{l,m}$ is the projector operator given by $\protect{\textsl{P}_{l,m}=\ket{l_s,m_i}\bra{l_s,m_i}}$. If we work out this expression we will see that
\begin{eqnarray}
C_{l,m} &= & \eta N\rho_{l,m;l,m},
\end{eqnarray} and, therefore it is clear that, after doing the normalization
of the coincidence counts seen in this part of the experiment, one will be
able to determine the diagonal elements of the spatial qubits density operator
$\rho$  \footnote{Here is considered the usually accepted fair sampling
assumption where the ensemble observed in the experiment ($\eta \times N$ twin
photons) is considered to be a fair sample of the total ensemble of $\rho$
($N$ twin photons transmitted by the double slits).}.

To obtain four more MLI equations one should now introduce the detection double slits in the $z$-plane, with their slits at the transverse positions $x^{\mu}_{j,0}=0$ and $x^{\mu'}_{j,1}=\frac{\alpha \pi}{2d}$ (again, $\alpha =(z-z_{A})/k_p$), and record the coincidence rates with the detectors $D_j$'s behind these slits [See Fig.~\ref{Fig2}(b)]. The new coincidences rates are now written in terms of the projectors of Eq.~(\ref{Pz}) and are given by
\begin{eqnarray}
C_{x^{\mu}_{0},x^{\mu}_{0}} &= & \left(\frac{2b}{L}\right)^2\eta N Tr(\rho \textsl{P}_{x^{\mu}_{0},x^{\mu}_{0}}), \nonumber \\
C_{x^{\mu}_{0},x^{\mu'}_{1}} &= & \left(\frac{2b}{L}\right)^2\eta N Tr(\rho \textsl{P}_{x^{\mu}_{0},x^{\mu'}_{1}}), \nonumber \\
C_{x^{\mu'}_{1},x^{\mu}_{0}} &= & \left(\frac{2b}{L}\right)^2\eta N Tr(\rho \textsl{P}_{x^{\mu'}_{1},x^{\mu}_{0}}), \nonumber \\
C_{x^{\mu'}_{1},x^{\mu'}_{1}} &= & \left(\frac{2b}{L}\right)^2\eta N Tr(\rho \textsl{P}_{x^{\mu'}_{1},x^{\mu'}_{1}}),
\label{Czz}
\end{eqnarray}
where $L$ is the transverse size of the idler and signal diffraction patterns (caused by the apertures $A_j$'s) at the $z$-plane\footnote{This is, of course, an approximation where we assume that the total length of the double slit diffraction pattern, $L$, is defined by the transverse length of the three principal peaks of diffraction. However, we can say that Eq.~(\ref{Czz}) is a good approximation for the values of the coincidence rates considered since most of the intensity of the light scattered by a double slit lies in this $L$ region.}.
Therefore, it is clear that the factor $\frac{2b}{L}$ is a normalization constant which accounts for the effect of one photon diffraction, which causes a reduction of $\frac{2b}{L}$ in the detector's efficiency that is being used to record this photon at the $z$-plane. From these measurements one will be able to determine the imaginary and real parts of some of the
matrix elements of $\rho$. In Table~\ref{tab1}, we give a summary showing which are the components of $\rho$ that are determined from the measurements considered in this protocol. There is also a comparison with the measurements done in the QTR of two polarized qubits \cite{James}. In \ref{A2} we give the expressions which relate the elements of $\rho$ and all the coincidence rates considered in this protocol.

Let us now consider the detection of the signal photons behind its detection double slit at the $z$-plane and the detection of the idler photons behind the slits of the double slit $A_i$ (at the $z_A$-plane) as it is schematically shown in Fig.~\ref{Fig2}(c). The four possible coincidence rates obtained, again by permuting the detectors' positions behind the slits of these double slits, are now written as
\begin{eqnarray}
C_{x^{\mu}_{0},m} &= & \frac{2b}{L}\eta N Tr(\rho \textsl{P}_{x^{\mu}_{0},m}), \nonumber \\
C_{x^{\mu'}_{1},m} &= & \frac{2b}{L}\eta N Tr(\rho \textsl{P}_{x^{\mu'}_{1},m}),
\end{eqnarray}
with again $m=+$ ($m=-$) for the detection of the idler photon at the upper (lower) slit of $A_i$. The projector $\textsl{P}_{x^{\mu}_{k},m}$ is defined by
\begin{equation}
\textsl{P}_{x^{\mu}_{k},m} = \ket{f(x^{\mu}_{s,k}),m}\bra{f(x^{\mu}_{s,k}),m}.
\label{Pzaz}
\end{equation}
These other four equations are MLI equations \footnote{Here it should be clear that we are always considering that the MLI equations constructed are linearly independent in the system of sixteen equations that we want to construct. The reader can check this by calculating Eq.~(\ref{rel}), explicitly, for the values $x^{\mu}_{j,0}=0$ and $x^{\mu'}_{j,1}=
\frac{\alpha \pi}{2d}$ used.} and they allow the determination of new elements of $\rho$ as  is shown in Table~\ref{tab1}.

The last four MLI equations are obtained when the signal photons are detected behind the slits of $A_s$ and the idler photons behind the slits $x^{\mu}_{i,0}=0$ and $x^{\mu'}_{i,1}=\frac{\alpha \pi}{2d}$ at the transverse $z$-plane. The projectors which represent these four coincidence measurements are given by
\begin{equation}
\textsl{P}_{l,x^{\mu}_{k}} = \ket{l,f(x^{\mu}_{i,k})}\bra{l,f(x^{\mu}_{i,k})},
\label{Pzza}
\end{equation}
and the coincidence rates are
\begin{eqnarray}
C_{l,x^{\mu}_{0}} &= & \frac{2b}{L}\eta N Tr(\rho \textsl{P}_{l,x^{\mu}_{0}}), \nonumber \\
C_{l,x^{\mu'}_{1}} &= & \frac{2b}{L}\eta N Tr(\rho \textsl{P}_{l,x^{\mu'}_{1}}),
\label{czaz}
\end{eqnarray} where $l=+$ ($l=-$) holds for the detection of the signal photon at the upper (lower) slit of $A_s$. The elements of $\rho$ that can be determined from these last measurements are also shown in Table~\ref{tab1}.

So, after performing the sixteen measurements described above, one will have reconstructed the density operator $\rho$ of the two spatial qubits state generated in the experiment considered. It is worth mentioning that the protocol presented here can also be applied to any type of two spatial qubits state, including the mixed states recently reported in the literature \cite{GLima3}.

\begin{center}
\begin{table}
\begin{tabular}{c|c|c|c|c} \hline \hline
\multicolumn{5}{c}{\textbf{Reconstructing $\rho$}} \\
\hline \hline
Measur. & Pol. proj. & Coefficients &Spatial proj. & Coefficients \\ \cline{1-5}
1 & $\left\vert HH\right\rangle \left\langle HH\right\vert $ & $\rho_{H,H;H,H} $ & $\textsl{P}_{+,+}$ & $\rho _{+,+;+,+}$ \\
2 & $\left\vert HV\right\rangle \left\langle HV\right\vert $ & $\rho_{H,V;H,V} $ & $\textsl{P}_{+,-}$ & $\rho _{+,-;+,-}$ \\
3 & $\left\vert VH\right\rangle \left\langle VH\right\vert $ & $\rho_{V,H;V,H} $ & $\textsl{P}_{-,+}$ & $\rho _{-,+;-,+}$ \\
4 & $\left\vert VV\right\rangle \left\langle VV\right\vert $ & $\rho_{V,V;V,V} $ & $\textsl{P}_{-,-}$ & $\rho _{-,-;-,-}$ \\
5 & $\left\vert HD\right\rangle \left\langle HD\right\vert $ & $\mathbb{R}\left( \rho _{H,H;HV}\right) $ & $\textsl{P}_{+,x^{\mu}_{0}}$ & $\mathbb{R}(\rho _{+,+;+,-})$ \\
6 & $\left\vert HL\right\rangle \left\langle HL\right\vert $ & $\mathbb{I}\left( \rho _{H,H;H,V}\right) $ & $\textsl{P}_{+,x^{\mu'}_{1}}$ & $\mathbb{I}(\rho_{+,+;+,-})$\\
7 & $\left\vert VD\right\rangle \left\langle VD\right\vert $ & $\mathbb{R}\left( \rho _{V,V;V,H}\right) $ & $\textsl{P}_{-,x^{\mu}_{0}}$ & $\mathbb{R}(\rho _{-,-;-,+})$ \\
8 & $\left\vert VL\right\rangle \left\langle VL\right\vert $ & $\mathbb{I}\left( \rho _{V,V;V,H}\right) $ & $\textsl{P}_{-,x^{\mu'}_{1}}$ & $\mathbb{I}(\rho_{-,-;-,+}) $ \\
9 & $\left\vert DH\right\rangle \left\langle DH\right\vert $ & $\mathbb{R}\left( \rho _{H,H;V,H}\right) $ & $\textsl{P}_{x^{\mu}_{0},+}$ & $\mathbb{R}(\rho _{+,+;-,+})$ \\
10 & $\left\vert LH\right\rangle \left\langle LH\right\vert $ & $\mathbb{I}\left( \rho _{H,H;V,H}\right) $ & $\textsl{P}_{x^{\mu'}_{1},+}$ & $\mathbb{I}(\rho_{+,+;-,+}) $ \\
11 & $\left\vert DD\right\rangle \left\langle DD\right\vert $ & $\mathbb{R}\left( \rho _{H,H;V,V}\right) ,\mathbb{R}\left( \rho _{H,V;V,H}\right) $ &$\textsl{P}_{x^{\mu}_{0},x^{\mu}_{0}}$ & $\mathbb{R}\left( \rho _{+,+;-,-}\right) ,\mathbb{R}\left( \rho_{+,-;-,+}\right) $ \\
12 & $\left\vert DL\right\rangle \left\langle DL\right\vert $ & $\mathbb{I}\left( \rho _{H,H;V,V}\right) ,\mathbb{I}\left( \rho _{H,V;V,H}\right) $ & $\textsl{P}_{x^{\mu}_{0},x^{\mu'}_{1}}$ & $\mathbb{I}\left( \rho _{+,+;-,-}\right) ,\mathbb{I}\left(\rho _{+,-;-,+}\right) $ \\
13 & $\left\vert DV\right\rangle \left\langle DV\right\vert $ & $\mathbb{R}\left( \rho _{V,V;H,V}\right) $ & $\textsl{P}_{x^{\mu}_{0},-}$ & $\mathbb{R}(\rho _{-,-;+,-})$ \\
14 & $\left\vert LV\right\rangle \left\langle LV\right\vert $ & $\mathbb{I}\left( \rho _{V,V;H,V}\right) $ & $\textsl{P}_{x^{\mu'}_{1},-}$ &$\mathbb{I}(\rho_{-,-;+,-}) $ \\
15 & $\left\vert LD\right\rangle \left\langle LD\right\vert $ & $\mathbb{I}\left( \rho _{H,H;V,V}\right) ,\mathbb{I}\left( \rho _{H,V;V,H}\right) $ &$\textsl{P}_{x^{\mu'}_{1},x^{\mu}_{0}}$ & $\mathbb{I}\left( \rho _{+,+;-,-}\right) ,\mathbb{I}\left(\rho _{+-,-+}\right) $ \\
16 & $\left\vert LL\right\rangle \left\langle LL\right\vert $ & $\mathbb{R}\left( \rho _{H,H;V,V}\right) ,\mathbb{R}\left( \rho _{H,V;V,H}\right) $ &$\textsl{P}_{x^{\mu'}_{1},x^{\mu'}_{1}}$ & $\mathbb{R}\left( \rho _{+,+;-,-}\right) ,\mathbb{R}\left( \rho _{+,-;-,+}\right) $%
\label{tab1}
\end{tabular}
\caption{Summary of the spatial measurements needed to do the quantum
tomography of two spatial qubits. There is also a comparison with the
measurements used in the QTR of polarized qubits. $\ket{H}$, $\ket{V}$,
$\ket{D}$ and $\ket{L}$ are the kets representing polarized qubits with
horizontal, vertical, diagonal and left-circular polarization, respectively.
The indices $\mathbb{R}$ and $\mathbb{I}$ stand for the real and imaginary parts,
respectively.}
\label{tab1}
\end{table}
\end{center}

Here, we give a brief description of an experimental setup that can be used for implementing this
reconstruction scheme. This description is based on already known parameters from our experiments. First, to generate spatial qubits (qudits) we use two identical double (multi) slits aligned in the direction of the signal and idler beams, which are positioned
at a distance $z_a=200$~mm from the Type II BBO crystal. The slits's width is $0.1$~mm
with a separation of $0.25$~ mm. The photons transmitted through the double-slits are detected
in coincidence. In front of the detectors there are interference filters, with very small bandwidths (with a typical FWHM of $1$ nm) and centered at twice the pump beam wavelength, which will select the frequency of the generated signal and idler photons \cite{Monken,Edu1}. For reconstruction purposes identical single slits at the detection plane can be used at position $z-z_a=600$~mm. At this distance, the fourth order interference pattern
has a width of the order of $8$~mm. The detection slit has a width of $0.1$~mm.
This slit can be easily installed in front of single photon detector modules.
If a Kripton laser, with an average power of $200$~mW is used as a pump beam, a maximum number of $200$ coincidence counts in
$500$~s are expected for (b) configuration in Fig. 2. For this experimental setup
we have $x^{\mu}_{0}=0$ and $x^{\mu'}_{1}=0.496$~mm. These measurement positions are within the
fourth order interference pattern region and they can be well resolved. In case of the measurements
of type (a) and (c) $200$ coincidence counts are expected for measurement times of
$20$~s and $100$~s, respectively. Furthermore, there are not requirement
for numerical aperture of the detectors.

\subsection{Maximum likelihood estimation}

The technique of QTR is based on a linear inversion of the measured data as was shown above for the two spatial qubits states. Thus, it is dependent on any experimental errors that may occur while recording the data. They can appear as a consequence of experimental noise or misalignment and the reconstructed state is, in general, only a reasonable approximation of the real quantum state. The density matrices obtained may have properties that are not fully compatible with a quantum state.

To generate only possible density matrixes there is an alternative that has been used during the state determination. It is the numerical technique called maximum likelihood estimation \cite{Hradil}. However, even though it generates only possible density matrixes, it has the drawback of enhancing the uncertainty in the state estimation.

Once the two qubit density operator is reconstructed with the technique proposed here, one can apply the matrix obtained to this numerical optimization. In Ref \cite{James} a practical approach for doing this is extensively discussed and it can also be applied for two spatial qubits systems.

\section{Conclusion}

In conclusion, we have theoretically shown that one can use the technique of quantum tomography to reconstruct states composed of two spatial qubits. Even though we have considered the state determination only for this special case, we believe that it can be generalized and performed in a similar way for spatial states of more dimensions. The importance of this work comes from the possibility of using the spatial qubits for quantum communications protocols where the ability to characterize may be necessary. The main advantage that we can envisage at this stage is the possibility of using transverse correlations of the down-converted photons for encoding quantum information in qudits
instead of qubits. Besides, even in the case of qubits we see some other important advantages.
For instance, when entangled polarization qubits are propagated through optical fibers they suffer decoherence because of the depolarization effect of the fibers. The birrefrigence of the fiber gradually destroys the entanglement of the entangled polarized qubits. The entanglement of the time-bin entangled qubits also suffers of severe decoherence in optical fiber due to the chromatic
dispersion effect. However, when the spatial qubits are used for sending information thought optical fibers, both depolarization and chromatic dispersion does not affect, directly, their quantum correlations and thus the entanglement of the spatial qubits seems to be more robust against the decoherence effects of optical fibers. We have already started this line of research in our group to demonstrate this idea.

\ack

The authors would like to express their gratitude to Marcelo T. Cunha for having called
their attention to this problem and initiating the discussions which culminated in this
work. G. Lima, L. Neves and S. P\'adua were supported by CAPES, CNPq, FAPEMIG and
Mil\^enio Informa\c{c}\~ao Qu\^antica. C. Saavedra and A. Delgado were supported by
Grants Nos. FONDECYT 1061461 and Milenio ICM P06-67F. F. Torres was supported by MECESUP
UCO0209. This work is part of the international cooperation agreement CNPq-CONICYT
491097/2005-0.

\appendix
\section{Calculating the state $\ket{\Phi}$}
\label{A1}

As we showed in Section \ref{SECPSI}, the state of the down-converted photons when they are transmitted through double slits, placed at the $z_A$-plane, is given by Eq.~(\ref{estqubit}). Now, if we consider the free space propagation of the twin photons from these apertures to another transverse $z$-plane, we will obtain a new form for their state \cite{GLima2}:
\begin{eqnarray}
\left|\Psi\right\rangle_{z} &\propto &\int dq_{s}\int dq_{i} \ce{G} \left(q_{i},q_{s}\right)	
\left|1q_{i}\right\rangle	 \left|1q_{s}\right\rangle,
\end{eqnarray}
where the new two-photon amplitude $\ce{G} \left(q_{i},q_{s}\right)$ is
\begin{eqnarray}
\ce{G}\left(q_{s},q_{i}\right)	&=& \frac{a}{\pi}e^{-i\alpha q_{i}^{2}}e^{-i\alpha q_{s}^{2}}
\sinc\left(q_{i}a\right) \sinc\left(q_{s}a\right)
\left( W_{+,+}e^{-i\left(q_{i}+q_{s}\right)d}
+W_{-,-}e^{ i\left(q_{i}+q_{s}\right)d}\right)
\nonumber \\ &+ &
\frac{a}{\pi}e^{-i\alpha q_{i}^{2}}e^{-i\alpha q_{s}^{2}}\sinc\left(q_{i}a\right)
\sinc\left(q_{s}a\right)\left(W_{+,-}e^{-i\left(q_{i}-q_{s}\right)d}
+W_{-,+}e^{ i\left(q_{i}-q_{s}\right)d}
\right), \nonumber \\
\end{eqnarray}
again, $\alpha=\frac{(z-z_{A})}{k_p}$.

Now we consider the determination of the twin photon state after they are transmitted through the detection double slits considered in our protocol and which are placed now at this transverse $z$-plane. The transverse position for the slits of these double slits are $x_{j,0}^{\mu}$ and $x_{j,1}^{\mu}$. To calculate this state we first assume a general form for it
\begin{eqnarray}
		\left|\Phi\right\rangle \propto \int dq_{s}\int dq_{i} \ce{D}
\left(q_{s},q_{i}\right)	\left|1q_{s}\right\rangle	 \left|1q_{i}\right\rangle
		\label{estado final 1}.
\end{eqnarray}
The two-photon amplitude $\ce{D} \left(q_{s},q_{i}\right)$ can be written in terms of the convolution
\begin{eqnarray}
\ce{D}(q_{s},q_{i}) &=\int\!\! dq'_{s}\!\!\int\!\! dq'_{i}
\ce{G}(q'_{s},q'_{i}) T_s(q'_{s}-q_{s})T_i(q'_{i}-q_{i}), 
\label{D1}\end{eqnarray}
where $T_j$'s are the Fourier transform of the transmission function of the detection double slits and are given by
\begin{eqnarray}
T(q_{j}) &=& 2b\!\left[e^{iq_{j}x_{j,0}^{\mu}}
\sinc\left(q_{j}b\right)+e^{iq_{j}x_{j,1}^{\mu}}\sinc\left(q_{j}b\right)\right], 
\label{T}
\end{eqnarray}
and by replacing Eq.~(\ref{T}) into Eq.~(\ref{D1}) we obtain
\begin{eqnarray}
	\ce{D}\left(q_{s},q_{i}\right) &= & \int dq'_{s}
\int dq'_{i}e^{-i\alpha( q_{i}^{'2}+ q_{s}^{'2})}
\sinc\left(q'_{i}a\right)
\sinc\left(q'_{s}a\right)
\sinc\left((q'_{i}-q_{i})b\right)
\nonumber\\& & \times \sinc\left((q'_{s}-q_{s})b\right) \sum_{u,v=\pm}W_{u,v}e^{-\left(u q'_{i}+ v q'_{s}\right)}
\times \sum_{k,k'=0,1}e^{i(q'_{s}-q_{s})x_{s,k}^{\mu}}e^{i(q'_{i}-q_{i})x_{i,k'}^{\mu}}. \nonumber \\
\end{eqnarray}
By using the following solution
\begin{eqnarray}
	I_{\pm}(x_{j,k}^{\mu},q_{j})\!\!&=&\!\!\!\!\int e^{-i\alpha q_{j}^{'2}} e^{\pm i q_{j}' d} e^{i(q_{j}'-q_{j})x_{j,k}^{\mu}} \sinc\left(q_{j}'a\right)\sinc\left((q_{j}'-q_{j})b\right)dq_{j}'\nonumber\\
	&\approx&	\!\!
e^{-iq_{j}x_{j,k}^{\mu}}e^{i\frac{\left(x_{j,k}^{\mu}\pm d\right)^{2}}{4\alpha}}
\sinc\left(\frac{\left(x_{j,k}^{\mu}\pm d\right)a}{2\alpha}\right)\sinc\left(\frac{x_{j,k}^{\mu}b}{2\alpha}+q_{j}b\right), \nonumber \\
\end{eqnarray}
and by defining the coefficients
\begin{eqnarray}
	r_{\pm}(x_{j,k}^{\mu})\equiv \sinc\left(\frac{\left(x_{j,k}^{\mu}\mp d\right)a}{2\alpha}\right)\exp\left(i\frac{(x_{j,k}^{\mu}\mp d)^{2}}{4\alpha}\right),
\end{eqnarray}
one can write the function $\ce{D}\left(q_{s},q_{i}\right)$ as
\begin{eqnarray}
	\ce{D}\left(q_{s},q_{i}\right)&=& \sum_{k,k'=0,1}\sum_{u,v=\pm}	W_{u,v}
	r_{u}(x_{s,k}^{\mu})r_{v}(x_{i,k'}^{\mu})e^{-iqx_{i,k'}^{\mu}} \sinc\left(\frac{x_{i,k'}^{\mu}b}{2\alpha}+qb\right) \nonumber \\ & & \times
e^{-iqx_{s,k}^{\mu}}\sinc\left(\frac{x_{s,k}^{\mu}b}{2\alpha}+qb\right).
\label{D2}
\end{eqnarray}
Now if we replace Eq.~(\ref{D2}) into Eq.~(\ref{estado final 1}) and define the new slit states
\begin{eqnarray}
	\left|f(x_{j,k}^{\mu})\right\rangle&\equiv&\sqrt{\frac{b}{\pi}} \int dq_{j}e^{-iq_{j}x_{j,k}^{\mu}}\nonumber\\
&&\times\sinc\left(\left[\frac{x_{j,k}^{\mu}}{2\alpha}+q_{j}\right]b\right)\left|1q_{j}\right\rangle,
\end{eqnarray}
we get that the state of the twin photons which were transmitted by the detection double slits is
\begin{eqnarray}
\ket{\Phi}&\propto& \sum_{k,k'=0,1}\sum_{u,v=\pm}	W_{u,v}
r_{u}(x_{s,k}^{\mu})r_{v}(x_{i,k'}^{\mu})\left|f(x_{s,k}^{\mu})\right\rangle\left|f(x_{i,k'}^{\mu})\right\rangle\\	 &=&\sum_{k,k'=0,1}B_{k,k'}\left|f(x_{s,k}^{\mu})\right\rangle\left|f(x_{i,k'}^{\mu})\right\rangle,
\end{eqnarray}
with
\begin{eqnarray}
	B_{k,k'}=\sum_{u,v=\pm} r_{u}(x_{s,k}^{\mu})r_{v}(x_{i,k'}^{\mu})W_{u,v}.
\end{eqnarray}

\section{The coefficients of $\rho$}
\label{A2}

Here we show explicitly the expressions for determining all the coefficients of
the density operator, $\rho$, by following the measurement scheme given in Section
\ref{Prot} [See Fig. \ref{Fig2}]). The first type of measurement is performed with
the detectors $D_i$ and $D_s$ right behind the slits of the double slits $A_i$
and $A_s$, respectively (See Fig. \ref{Fig2}(a)). As it was discussed above,
the diagonal coefficients are given by
\begin{equation}
\rho _{l,m;l,m}=\frac{1}{\eta N}C_{l,m},
\end{equation}
with the normalization that $\eta N = C_{+,+}+C_{+,-}+C_{-,+}+C_{-,-}$. We
remind that $C_{l,m}$ is the coincidence rate recorded with the detector $D_s$
behind the slit $l$ of its double slit and with the detector $D_i$ behind the slit $m$ of $A_i$.

Now, we consider the detection of the signal photon behind its detection double slit at
$z$-plane [See Fig. \ref{Fig2}(c)] while the other detector remains behind the slits
of the aperture $A_s$ (See Section~\ref{Prot} for more details). In this case the coincidences
rates are
\begin{eqnarray}
C_{x^{\mu}_{0},m} &= & \frac{2b}{L}\eta N Tr(\rho \textsl{P}_{x^{\mu}_{0},m}), \nonumber \\
C_{x^{\mu'}_{1},m} &= & \frac{2b}{L}\eta N Tr(\rho \textsl{P}_{x^{\mu'}_{1},m}),
\end{eqnarray}
and by considering the values of $x^{\mu}_{s,0}=0$ and $x^{\mu'}_{s,1}=\frac{\alpha \pi}{2d}$
we get
\begin{eqnarray}
\rho _{+,+;-,+} &=&\frac{1}{\eta N}\left( \chi \left(
C_{x_{0}^{\mu},+}-iC_{x_{1}^{\mu'},+}\right) +\frac{C_{+,+}+C_{-,+}}{2}\left(
-1+i\right) \right) =\rho _{-,+;+,+}^{\ast }, \nonumber \\ \\
\rho _{+,-;-,-} &=&\frac{1}{\eta N}\left( \chi \left(
C_{x_{0}^{\mu},-}-iC_{x_{1}^{\mu'},-}\right) +\frac{C_{+,-}+C_{-,-}}{2}\left(
-1+i\right) \right) =\rho _{-,-;+,-}^{\ast }, \nonumber \\
\end{eqnarray}
where $\chi =L/(2b)$.

Reversing the role of signal and idler detectors, by considering the coincidence rates of Eq.~(\ref{czaz}) in our protocol, we obtain
\begin{eqnarray}
\rho _{+,+;+,-} &=&\frac{1}{\eta N}\left( \chi \left(
C_{+,x_{0}^{\mu}}-iC_{+,x_{1}^{\mu'}}\right) +\frac{C_{+,+}+C_{+,-}}{2}\left(
-1+i\right) \right) =\rho _{+,-;+,+}^{\ast }, \nonumber \\ \\
\rho _{-,+;-,-} &=&\frac{1}{\eta N}\left( \chi \left(
C_{-,x_{0}^{\mu}}-iC_{-,x_{1}^{\mu'}}\right) +\frac{C_{-,+}+C_{-,-}}{2}\left(
-1+i\right) \right) =\rho _{-,-;-,+}^{\ast }. \nonumber \\
\end{eqnarray}

The last elements of $\rho$ to be calculated are the anti-diagonal coefficients.
To determine them we need to consider the measurements which involve the propagation
of both photons, i.e, the coincidences rates of Eq.~(\ref{Czz}). The anti-diagonal
elements are given in terms of these coincidences by
\begin{eqnarray}
\mathbb{R}\left( \rho _{+,+;-,-}\right)  =\frac{1}{\eta N}\left[ \chi
^{2}\left( C_{x_{0}^{\mu},x_{0}^{\mu}}-C_{x_{1}^{\mu'},x_{1}^{\mu'}}\right)\right]
\nonumber \\ \times \frac{1}{\eta N}\left[ -\frac{%
\chi \left(
C_{+,x_{0}^{\mu}}-C_{+,x_{1}^{\mu'}}+C_{-,x_{0}^{\mu}}-C_{-,x_{1}^{\mu'}}
+C_{x_{0}^{\mu},+}-C_{x_{1}^{\mu'},+}+C_{x_{0}^{\mu},-}-C_{x_{1}^{\mu'},-}\right)
}{2}\right],\nonumber \\
\end{eqnarray}

\begin{eqnarray}
\mathbb{R}\left( \rho _{+,-;-,+}\right) =\frac{1}{\eta N}\left[ \chi
^{2}\left( C_{x_{0}^{\mu},x_{0}^{\mu}}+C_{x_{1}^{\mu'},x_{1}^{\mu'}}\right) \right]
\nonumber \\ \times \frac{1}{\eta N} \left[-\frac{\chi \left(C_{+,x_{0}^{\mu}}
+C_{+,x_{1}^{\mu'}}+C_{-,x_{0}^{\mu}}+C_{-,x_{1}^{\mu'}}
+C_{x_{0}^{\mu},+}+C_{x_{1}^{\mu'},+}+C_{x_{0}^{\mu},-}+C_{x_{1}^{\mu'},-}\right)
}{2}\right] +\frac{1}{2}, \nonumber \\
\end{eqnarray}

\begin{eqnarray}
\mathbb{I}\left( \rho _{+,+;-,-}\right)  =\frac{1}{\eta N}\left[ -\chi
^{2}\left( C_{x_{0}^{\mu},x_{1}^{\mu'}}+C_{x_{1}^{\mu'},x_{0}^{\mu}}\right) \right]
\nonumber \\ \times \frac{1}{\eta N} \left[\frac{\chi \left(
-C_{+,x_{0}^{\mu}}+C_{+,x_{1}^{\mu'}}-C_{-,x_{0}^{\mu}}+C_{-,x_{1}^{\mu'}}
+C_{x_{0}^{\mu},+}-C_{x_{1}^{\mu'},+}+C_{x_{0}^{\mu},-}-C_{x_{1}^{\mu'},-}\right)
}{2}\right] +\frac{1}{2}, \nonumber \\
\end{eqnarray}

\begin{eqnarray}
\mathbb{I}\left( \rho _{+,-;-,+}\right)  =\frac{1}{\eta N}\left[ \chi
^{2}\left( C_{x_{0}^{\mu},x_{1}^{\mu'}}-C_{x_{1}^{\mu'},x_{0}^{\mu}}\right) \right]
\nonumber \\ \times \frac{1}{\eta N} \left[-\frac{%
\chi \left(
C_{+,x_{0}^{\mu}}+C_{+,x_{1}^{\mu'}}+C_{-,x_{0}^{\mu}}+C_{-,x_{1}^{\mu'}}
+C_{x_{0}^{\mu},+}+C_{x_{1}^{\mu'},+}+C_{x_{0}^{\mu},-}+C_{x_{1}^{\mu'},-}\right)
}{2}\right] + 1. \nonumber \\
\end{eqnarray}

\end{document}